\documentclass[aps,preprint,amsmath,amssymb]{revtex4-1}
\usepackage{graphicx}
\newcommand{\nn}{\nonumber}
\newcommand{\bd}{\begin{document}}
\newcommand{\ed}{\end{document}}
\newcommand{\bc}{\begin{center}}
\newcommand{\ec}{\end{center}}
\newcommand{\be}{\begin{eqnarray}}
\newcommand{\ee}{\end{eqnarray}}
\newcommand{\ba}{\begin{array}}
\newcommand{\ea}{\ed{array}}
\newcommand{\strich}[1]{#1  \! \! \slash}
\newcommand{\eqn}{\global\def\theequation}
\newcommand{\sw}{sin^2 \theta_W}
\newcommand{\fbd}{f_B}
\renewcommand{\thefootnote}{\alph{footnote}}
\newcommand{\se}{\section}
\newcommand{\sse}{\subsection}
\newcommand{\bi}{\bibitem}
\def\figcap{\section*{Figure Captions\markboth
     {FIGURECAPTIONS}{FIGURECAPTIONS}}\list
     {Figure \arabic{enumi}:\hfill}{\settowidth\labelwidth{Figure 999:}
     \leftmargin\labelwidth
     \advance\leftmargin\labelsep\usecounter{enumi}}}
\let\endfigcap\endlist \relax
\def\reflist{\section*{References\markboth
     {REFLIST}{REFLIST}}\list
     {[\arabic{enumi}]\hfill}{\settowidth\labelwidth{[999]}
     \leftmargin\labelwidth
     \advance\leftmargin\labelsep\usecounter{enumi}}}
\let\endreflist\endlist \relax

\def\Journal#1#2#3#4{{#1} {{\bf #2},} {#4} {(#3)}}
\def\NCA{Nuovo Cimento}
\def\NIM{Nucl. Instrum. Methods}
\def\NIMA{{Nucl. Instrum. Methods} A}
\def\NP{{Nucl. Phys.} }
\def\NPB{{Nucl. Phys.} B }
\def\NPA{{Nucl. Phys. A}}
\def\PLB{{Phys. Lett.}  B}
\def\PL{{Phys. Lett.}}
\def\PPSA{{Proc. Phys. Soc.} A}
\def\PRP{{ Phys. Rep.}}
\def\PRL{ Phys. Rev. Lett.}
\def\PR{{Phys. Rev.}}
\def\PRD{{Phys. Rev.} D}
\def\PRC{{Phys. Rev.} C}
\def\ZP{{Z. Phys.}}
\def\ZPC{{Z. Phys. C}}
\def\EPJ{{Eur. Phys. J.}}
\def\EPJC{{Eur. Phys. J.} C}
\def\ZPA{{Z. Phys.} A}
\def\MPL{{Mod. Phys. Lett.}}
\def\MPLA{{Mod. Phys. Lett.} A}
\def\CPC{Comput. Phys. Commun.}
\def\JHEP{{J. High Energy Phys.}}
\def\JPG{{J. Phys. G.}}
\def\SJNP{Sov. J. Nucl. Phys.}
\def\NCA{ Nuovo Cimento}
\def\NIM{ Nucl. Instrum. Methods}
\def\NIMA{{ Nucl. Instrum. Methods} A}
\def\NP{{ Nucl. Phys.}}
\def\ANP{{Adv. Nucl. Phys.}}
\def\CPC{{Comput. Phys. Commun.}}
\begin{document}
\title
{\Large {\bf Form factors of $\eta_{c}$ in light front quark model}
}
\author{
Chao-Qiang Geng$^{1,3,4}$\footnote{E-mail address: geng@phys.nthu.edu.tw}
and 
Chong-Chung Lih$^{2,3,4}$\footnote{E-mail address: 
cclih@phys.nthu.edu.tw}  
}
\affiliation{
$^1$College of Mathematics \& Physics, Chongqing University of Posts \& Telecommunications, Chongqing, 400065, China\\
$^2$Department of Optometry, Shu-Zen College of Medicine and Management,
Kaohsiung Hsien,Taiwan 452   \\
$^3$Physics Division, National Center for Theoretical Sciences, Hsinchu, Taiwan 300\\
$^4$Department of Physics, National Tsing Hua University, Hsinchu, Taiwan 300 
}
\date{\today}

\begin{abstract}
We study the form factors of  the $\eta_{c}$ meson in the light-front quark model. 
We explicitly  show that the transition form factor of  $\eta_c \to \gamma^* \gamma$ 
as a function of the momentum transfer is consistent with the experimental data 
by the BaBar collaboration, while the decay constant of  $\eta_c$ is found to be
$f_{\eta_{c}}=230.5^{+52.2}_{-61.0}$
 and $303.6^{+115.2}_{-116.4}$   MeV 
for $\eta_c\sim c\bar{c}$ by using  two  
$\eta_c \to \gamma \gamma$ decay widths of $5.3\pm 0.5$
and $7.2\pm2.1$ keV, given by  Particle Data Group and Lattice QCD calculation, respectively.
\end{abstract}


\maketitle %

\se{Introduction}

A neutral meson decaying into two photons  is the simplest exclusive process
since there is  only  one form factor  involved.
Experimental searches have been concentrated on
the transition form factors of $P \to \gamma^*\gamma$ $(P=\pi^0, \eta$ and $\eta')$, $F_{P\gamma}(Q^2)$,
 in terms  of the momentum transfer $Q^2$.
In particular, the anomalous result for the pion transition form factor of  $F_{\pi\gamma}(Q^2)$
at the large $Q^2$ has been reported  by the BaBar collaboration~\cite{BaBarPi}
in comparison with the theoretical expectation~\cite{ThPi} as well as the recent data by the Belle 
collaboration~\cite{Belle}. In addition,
 the transition form factors of $\eta,\eta^\prime\to\gamma^*\gamma$, $i.e.$
$F_{\eta\gamma,\eta'\gamma}(Q^2)$, have also been measured by BaBar~\cite{BaBar1,BaBar2}
 in the regions up to 40 and 35 GeV$^2$,  respectively. 
These form factors as functions of  $Q^2$ 
allow us to not only extract information on the meson wave functions, but also check the pQCD predictions. 
Recently, we have studied  the transition form factors of $F_{P\gamma}(Q^2)$
at the large $Q^2$ in the light-front quark model (LFQM)~\cite{lfpi,Geng:2012qg}
and shown that our results can fit with all the  data, including those at the large $Q^2$ regions.

There is another interesting  form factor, which is related to the $\eta_{c}\to \gamma^*\gamma$ transition. 
The 
measurements on this form factor have been done by both L3~\cite{L3Etac} and
 BaBar~\cite{BaBarEtac} Collaborations based on the process of
$e^+ e^-\to e^+ e^-\gamma^* \gamma^*\to e^+ e^- \eta_c$ in the range of $Q^2$ from 2 to 50 GeV$^2$. 
Since $\eta_c$ is composed of two massive charm quarks, it is important to know
the behavior of the $\eta_{c}\to \gamma^*\gamma$
transition form factor at a high Q$^2$ momentum transfer to
examine the validity of the pQCD calculations in this heavy meson
as well as compare with those of the light 
pseudoscalars.
The transition form factor of $F_{\eta_c\gamma}(Q^2)$ has been  investigated by various 
QCD models~\cite{etac1}. 
In this paper, we study this form factor
for $Q^2$ up to 50 GeV$^2$ in the LFQM. 
We will also simultaneously evaluate the $\eta_c$ meson decay 
constant $f_{\eta_c}$. This form factor is
important for us to understand the pseudoscalar charmonium meson $\eta_c$.
Moreover, the precise knowledge of $f_{\eta_c}$ can  help us to examine
other related processes, such as $B\to \eta_cK$.

 By analogy with the $\pi^{0}\gamma$ and $\eta^{(')}\gamma$ 
cases, we use the light front approach based on the simple meson wave function 
structure of  $Q\bar{Q}$ ($Q=u,d,s,c$) pairs, constrained by 
the experimental measured decay 
width of $\eta_c\to \gamma\gamma$ as well as the mass of $\eta_c$.

This paper is organized as follows.  In Sec.~II, we present 
the transition form factors for $Q\bar{Q} \to \gamma^* \gamma$. 
In Sec.~III, we show our numerical results on the transition  form factor
of $\eta_{c}\to \gamma^*\gamma$
and  the decay constant of $\eta_{c}$. 
We give our conclusions in Sec.~IV.

\se{The form factors}

\begin{figure}[htbp]
\includegraphics*[width=2in,height=6in,angle=-90]{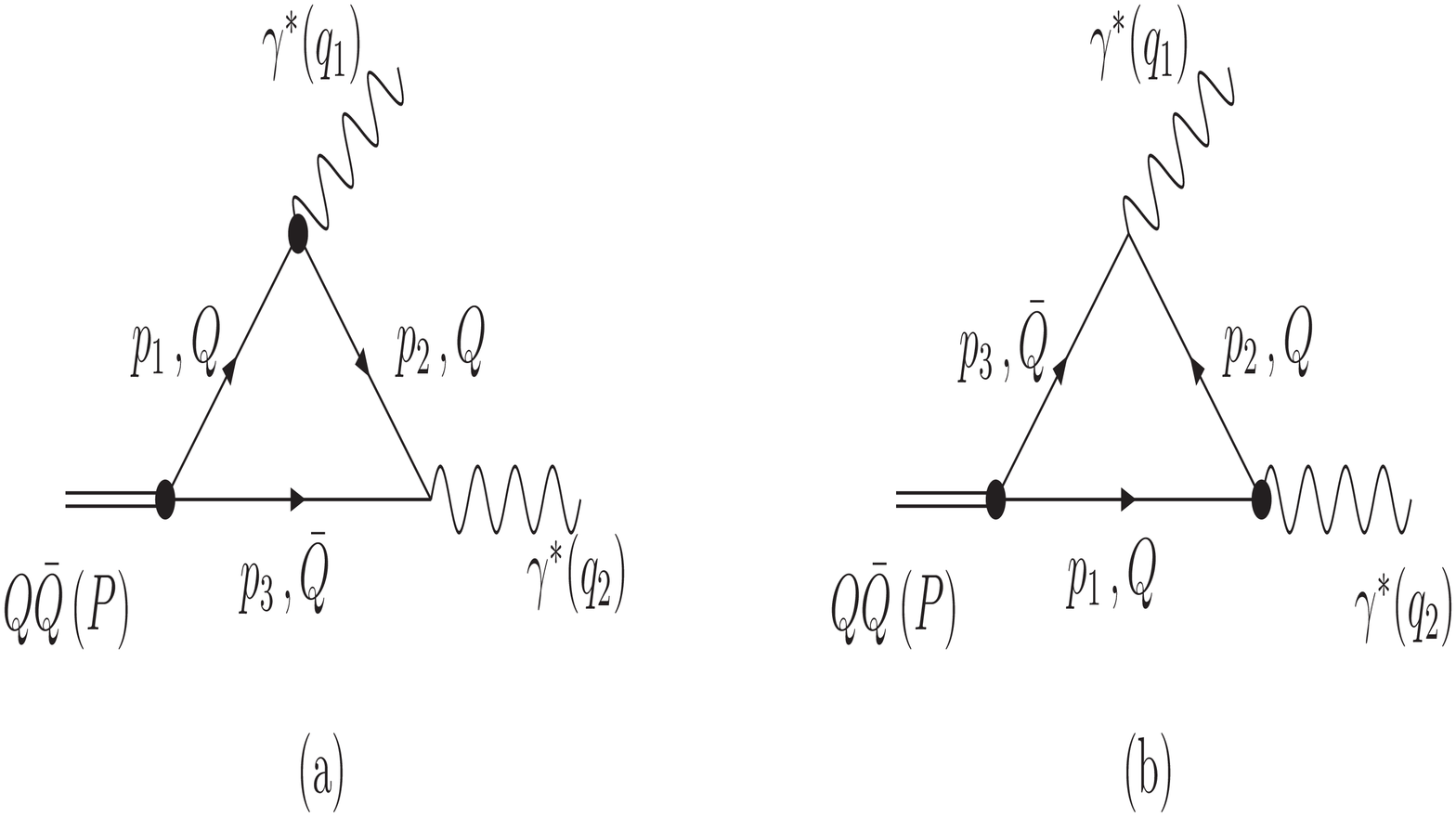}
\caption{  Loop diagrams contributing to $Q\bar{Q} \to \gamma^*\gamma^*$.}
\label{F1}
\end{figure}

To extract the transition form factor $F_{\eta_c\gamma}$ , we first write the
decay amplitude of $\eta_{c} \to \gamma^*\gamma^*$ as~\cite{vex3}
\be 
A(\eta_{c}\to\gamma^*(q_1,\epsilon_1)~\gamma^*(q_2,\epsilon_2))
=ie^{2}F_{\eta_{c}\gamma}(q^2_1,q^2_2)~\varepsilon_{\mu\nu\rho\sigma}~\epsilon^\mu_1
 ~\epsilon^\nu_2 ~q^\rho_1 ~q^\sigma_2\,, \label{def}
\ee
where $F_{\eta_{c}\gamma}(q^2_1,q^2_2)$ 
is a symmetric function under the interchange of $q^2_1$ and $q^2_2$. 
The light front approach \cite{lf1,lf6} provides a framework for the relativistic quark model in which a
consistent and relativistic treatment of quark spins and the center-of-mass motion can be carried out. 
In this framework,
we consider the meson wave function as a combination of the $Q\bar{Q}$ Fock states. 
In the quark-flavor mixing scheme, the state of $\eta_c$ can be parameterized as~\cite{etacPRD}
\be
|\eta_c\rangle &=&
 -\theta_c\sin(\phi-\theta_y)\,|\psi_{q}\rangle-
\theta_c\cos(\phi-\theta_y)\,|\psi_{s}\rangle+ |\psi_{c}\rangle\,,
\label{etac}
\ee
where $\theta_{c,y}$ and $\phi$ are the mixing angles and
 $|\psi_{q}\rangle=\frac{1}{\sqrt{2}}|u\bar{u}+d\bar{d}\rangle$, 
$|\psi_{s}\rangle=|s\bar{s}\rangle$ and $|\psi_{c}\rangle=|c\bar{c}\rangle$
in the quark flavor basis.
%
Consequently,  $F_{\eta_{c}}(q^2_1,q^2_2)$ can be found by summing up the relevant $Q\bar{Q}$ Fock states. 
From the quark-$Q\bar{Q}$ meson loops shown in Fig.~\ref{F1}, 
we get~\cite{lfpi,Geng:2012qg}
\be
A(Q\bar{Q}\to \gamma^*(q_1)~\gamma^*(q_2)) &=&
e_{Q}e_{\bar{Q}} N_{c}\int {d^4 p_3 \over{(2 \pi)^4}}
\Lambda_{Q\bar{Q}}\Bigg\{{\rm Tr}\Bigg[\gamma_5
        {i(-\not{\! p_3}+m_{\bar{Q}})\over{p_3^2-m^2_{\bar{Q}}+i\epsilon}}\not{\! \epsilon_2}
       {i(\not{\!p_2}+m_Q)\over{p_2^2-m^2_Q+i\epsilon}} \nn \\
&&\times \not{\! \epsilon_1}
        {i(\not{\! p_1}+m_Q)\over{p_1^2-m^2_Q+i\epsilon}}
\Bigg]+(\epsilon_1 \leftrightarrow \epsilon_2\,,\,q_1 \leftrightarrow q_2) \Bigg\}
 \nn \\
&&+(\,p_{1(3)} \leftrightarrow p_{3(1)}\,,\, m_{Q} \leftrightarrow
m_{\bar{Q}}) \,,
\label{matrix}
\ee
where $N_c$ is the number of colors, $\Lambda_{Q\bar{Q}}$ is a vertex function  related to the
momentum distribution amplitude of the $Q\bar{Q}$ Fock state, and $m_Q=m_{\bar{Q}}$ is the Q-quark mass.
In the LFQM, the amplitude  can be solved in principle by 
the light-front QCD bound state equation~\cite{lf1}. 
Here, however,  we chose
 phenomenological Gaussian type of the amplitude~\cite{lf6,Geng:2012qg,lfpi}.
 Similar to the procedures in Refs.~~\cite{Geng:2012qg,lfpi},
 after integrating over $p_3^-$ and calculating the trace in Eq.~(\ref{matrix}),
 we obtain
the form factor $F_{Q\bar{Q}}(q^2_1,q^2_2)$ in Eq.~(\ref{def})  to be: 
\be
F_{Q\bar{Q}}(q_{1}^{2},q_{2}^{2}) &=&-{16\over 9} \sqrt{N_{c}\over 3}
        \int \frac{dx\,d^{2}k_{\bot }}{2\left( 2\pi \right) ^{3}}\Phi
        \left( x,k_{\bot }^{2}\right) {1\over 1-x}
\frac{m_{Q}+(1-x)m_{Q} k_{\bot}^{2}\Theta}{x(1-x)q_{2}^{2}-m_{Q}^{2}-k_{\bot }^{2}}
       +(q_2 \leftrightarrow q_1)  \,,~~
\label{fffv}
\ee
with
\be
\Phi (x,k_{\bot}^2) = \sqrt{{\frac{x(1-x) }{2 M_0^2}}} \phi_{Q\bar{Q}}(x, k_\perp)\,,\;\;
\Theta = {\frac{1}{\Phi(x,k_{\bot}^2) }} {\frac{d\Phi(x,k_{\bot}^{2})}{%
dk_{\bot}^2}} \, , 
\label{res}
\ee
where
\be
\label{res}
M_0^2 &=&{ m_{Q}^2+k_\bot^2\over x}+{ m_{Q}^2+k_\bot^2\over
1-x}\, ,
\nonumber\\
\phi_{Q\bar{Q}}(x, k_\perp) &=& N \sqrt{{\frac{dk_{z}}{dx}}}\exp \left( -{\frac{\vec{k}^{2}}{%
2\omega_{Q\bar{Q}}^2}}\right)\,,
%
\label{n6}
\ee
with 
 $N = 4 ( \pi/\omega_{Q\bar{Q}}^2)^{3/4}$,  $\vec{k} =(k_{\bot}, k_z)$,  $k_z=(x+1/2)M_0$,
and $\omega_{Q\bar{Q}}$ being the parameter related to the physical size of a pseudoscalar state of $ Q\bar{Q}$
in the wave function.
If $q_1^2\equiv Q^2$ and $q_2^2=0$, $i.e.$ one of the photons is on its mass shell,
from Eq.~(\ref{fffv}) we derive
\be
F_{Q\bar{Q}} (Q^2)&\equiv& F_{Q\bar{Q}} (Q^2,0) 
\nonumber\\
&=&-{16\over 9} \sqrt{N_{c}\over 3}
       \int \frac{dx\,d^{2}k_{\bot }}{2\left( 2\pi \right) ^{3}} {\Phi
        \left( x,k_{\bot }^{2}\right)\over 1-x}
\left\{ \frac{m_{Q}+(1-z)m_{Q} k_{\bot}^{2}\Theta}{x(1-x)Q^{2}-m_{Q}^{2}-k_{\bot }^{2}}
-\frac{m_{Q}}{m_{Q}^{2}+k_{\bot }^{2}} \right\} .
\label{realff}
\ee
Consequently, the transition form factor of $F_{\eta_{c}\gamma}$ can be found by
\be
F_{\eta_{c}\gamma}(Q^2) =  -\theta_c\sin(\phi-\theta_y) F_{q\bar{q}} (Q^2)-
\theta_c\cos(\phi-\theta_y)F_{s\bar{s}}(Q^2)  + F_{c\bar{c}}(Q^2) \,,
\label{realff1}
\ee
with $F_{Q\bar{Q}} (Q^2)$ ($Q=q,s,c$) given in Eq.~(\ref{realff}).
The decay width of $\eta_c\to \gamma\gamma$ is related to the form factor of $F_{\eta_c\gamma}(Q^2= 0)$ by
\be
\label{DWidth}
\Gamma_{\eta_{c} \to 2\gamma}&=& \frac{(4\pi \alpha)^{2}
}{64\pi} m_{\eta_{c}}^{3} |F_{\eta_{c}\gamma}(0)|^2 \,.
\ee
 Based on the experimental data of $\Gamma_{\eta_{c} \to 2\gamma}$, from Eqs.~(\ref{realff})-(\ref{DWidth})
 we can extract the free parameter in the meson wave function.

The decay constant of the $\eta_c$ meson is defined by the matrix element
\be
\langle 0|\bar{q} \gamma_{\mu}\gamma_{5}q|\eta_{c}(P)\rangle=i f_{\eta_{c}}P_{\mu}.
\ee
Combining the above with  Eq.~(\ref{etac}), we obtain the decay constant of $\eta_c$ to be
\be
\label{fq0}
f_{\eta_c} &=& 
 -\theta_c\sin(\phi-\theta_y) f_{q\bar{q}}-
\theta_c\cos(\phi-\theta_y)f_{s\bar{s}}+ f_{c\bar{c}}\,,
\ee
where the explicit expressions of $f_{Q\bar{Q}}$ $(Q=q,s,c)$ are given by~\cite{Geng:2012qg,fp}
\be
f_{Q\bar{Q}}&=&\,4{\sqrt{N_c}\over\sqrt{2}}\int {dx\,d^2k_\perp\over 2(2\pi)^3}\,\phi_{Q\bar{Q}}(x,
k_\perp)\,{m_Q\over\sqrt{m_Q^{2}+k_\perp^2}}\,.
\label{fq}
\ee

\se{Numerical Result}

To numerically calculate the transition form factor of $F_{\eta_c\gamma}$
in Eq.~(\ref{realff1}), we need to
specify the parameters in the meson wave functions, in particular the meson scale
parameters of $\omega_{Q\bar{Q}}$ in Eq.~(\ref{res}).
 From Eq.~(\ref{DWidth}),  $F_{\eta_c\gamma}(0)$ can be fixed by
 $\Gamma_{\eta_{c} \to 2\gamma}$, which can be used to determine
 the value of $\omega_{c\bar{c}}$.
 Note that the scale parameters of $\omega_{q\bar{q},s\bar{s}}$ can be found in Refs.~\cite{lfpi,Geng:2012qg}.
  In our calculations, we use $m_{\eta_{c}}=2981.0\pm1.1$ MeV~\cite{pdg} and 
   two inputs for the decay width of $\eta_c\to\gamma\gamma$:
 \be
 \label{2inputs}
 5.3\pm0.5\,\,\rm{keV} \;\; ({\rm I})\;\;\;  &{\rm and}& \;\;\;7.2\pm2.1\,\,\rm{keV}\;\; ({\rm II})\,,
 \ee
  given by Particle Data Group (PDG)~\cite{pdg} and  Lattice QCD prediction~\cite{etacdec2},
  which lead to 
 \be
|F_{\eta_c\gamma}(0)|=0.069\pm0.003\,\,\rm{GeV^{-1}}\;\; ({\rm I})\;\;\; &{\rm and}&  \;\;\; 0.081\pm0.011\,\,\rm{GeV^{-1}}\;\; ({\rm II})\,,
\ee
 respectively. 
 The mixing angles have beed studied in Ref.~\cite{etacPRD}. To illustrate the effects of the mixings, we take 
 \be
 \label{Mixing}
 (a)~~(\theta_c,\theta_y,\phi)&=& (-1.0^{\circ},-21.2^{\circ},39.3^{\circ})\,,
 \nonumber\\
 (b)~~(\theta_c,\theta_y,\phi)&=& (-0.9^{\circ},-21.2^{\circ},42.0^{\circ})\,,
 \nonumber\\
(c)~~(\theta_c,\theta_y,\phi)&=& ( 0^{\circ},-,-)\,.
 \ee
 We note that the case (a)  corresponds the center values used in Ref.~\cite{etacPRD}, while the case (c) represents that $\eta_c$ is
 a pure $c\bar{c}$ state.
We also note that there is no other free parameter to adjust the light front wave function of $c\bar{c}$. 
Now, we can fit the parameter of $\omega_{c\bar{c}}$ from Eqs.~(\ref{realff}) and  (\ref{realff1})
with $q^2=0$ and a given charm quark mass. In the $\overline{MS}$ scheme, one has that
$\bar{m}_c(\bar{m}_c)=1.29^{+0.05}_{-0.11}$ GeV~\cite{pdg}. However, 
at the higher order in QCD, the center value is enhanced, 
while the range of the pole charm quark mass  is even border~\cite{Alekhin:2012vu}.
In our numerical calculation, $m_c$ is a free input parameter.

\begin{figure}[htbp]
\includegraphics*[width=4in,height=3in,angle=0]{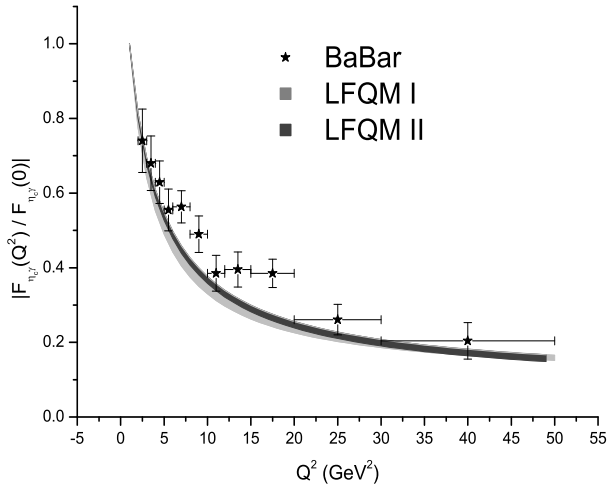}
\caption{ $|F_{\eta_c\gamma}(Q^2)/F_{\eta_c\gamma}(0)|$  
as a function of $Q^2$ in the LFQM, where the bands for LFQM I and II 
correspond to the calculations based on $\Gamma_{\eta_c\to\gamma\gamma}$ 
given by PDG~\cite{pdg} and Lattice QCD calculations~\cite{etacdec2},
respectively,
while the stars  represent the experimental data by the BaBar collaboration~\cite{BaBarEtac}.
}
\label{F2}
\end{figure}
In Fig.~\ref{F2}, 
we show 
the numerical results for $F_{\eta_{c}\gamma} (q^2)/F_{\eta_{c}\gamma}(0)$ in the LFQM for the first set of the mixing angles in Eq.~(\ref{Mixing}),
where I and II represent the two inputs of $\Gamma_{\eta_c\to\gamma\gamma}$ in Eq.~(\ref{2inputs}), respectively.
From the figure, we see that both  predictions in the LFQM I and II 
are consistent with the BaBar experimental data even though they are about 10$\%$ smaller
than the data points in the range (7.5$\sim$20) GeV.  We remark that our results do not change much for the other sets of the mixing angles in Eq.~(\ref{Mixing}).
As an illustration, we can also fit  the result of the LFQM I by
a double pole form
\be
\frac {F_{\eta_{c}\gamma} (Q^2)}{F_{\eta_{c}\gamma}(0)}=\frac{1} 
{1+(Q/\alpha)^2-(Q/\beta)^4}
\label{fit}
\ee
where  $\alpha=2.2$ and $\beta=4.7$ in GeV and we have ignored the errors for the input parameters. 

\begin{figure}[htbp]
\includegraphics*[width=4in,height=3in,angle=0]{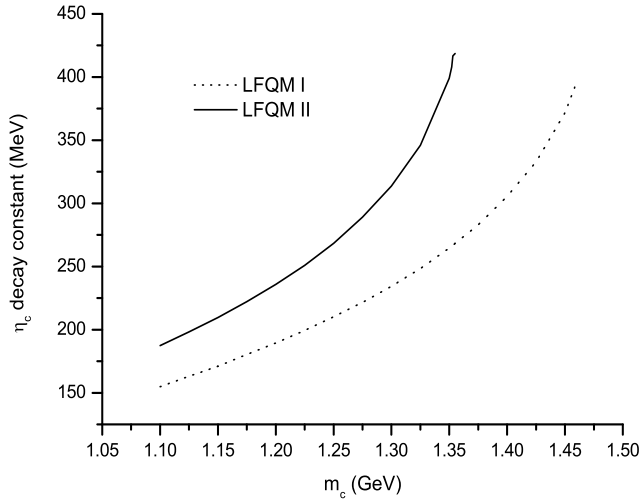}
\caption{ $\eta_c$ decay constant ($f_{\eta_c}$) 
as a function of $m_c$ in the LFQM without the mixing, i.e. $\theta_c=0$. 
}
\label{F3}
\end{figure}

 Apart from the transition form factor, from the fitted values of the meson scale
parameters of $\omega_{Q\bar{Q}}$ and Eq.~(\ref{fq}),
we can simultaneously determine the range of the $\eta_c$ decay constant $f_{\eta_c}$.
Our results on  $f_{\eta_c}$ with different sets of the mixing angles in Eq.~(\ref{Mixing}) and  the two inputs of (I) and (II) on 
$\Gamma_{\eta_{c}\to 2\gamma}$ in Eq.~(\ref{2inputs}) as well as 
the experimental data from the CLEO collaboration~\cite{etacdec1} and the 
Lattice QCD prediction~\cite{etacdec2} are summarized in Table.~\ref{T1}.
\begin{table}[htbp]
\caption{ The $\eta_{c}$ decay constant $f_{\eta_c}$  in the LFQM with three sets of  the mixing angles in Eq.~(\ref{Mixing}) and two input parameters in Eq.~(\ref{2inputs}), given
by PDG~\cite{pdg} and Lattice QCD~\cite{etacdec2},
respectively.
}
\vskip 0.2in
\label{T1}
\begin{tabular}{|c||c|c|} \hline
    & $\Gamma_{\eta_{c}\to 2\gamma}$ (keV) &  $f_{\eta_{c}}$ (MeV)
\\ \hline \hline
LFQM I & $5.3\pm0.5$~\cite{pdg}
&  (a) $194.0^{+33.3}_{-47.0}~$, (b)  $196.9^{+33.7}_{-44.0}$, (c) $230.5^{+52.2}_{-61.0}$
\\ \hline
LFQM II & $7.2\pm2.1$~\cite{etacdec2}
&  (a) $243.6^{+127.5}_{-84.4}$, (b) $249.2^{+143.2}_{-84.4}$, (c) $303.6^{+115.2}_{-116.4}$ 
\\ \hline
Lattice QCQ~\cite{etacdec2} & $7.2\pm2.1$~\cite{etacdec2}
&  $394.7\pm2.4$ 
\\ \hline
CLEO~\cite{etacdec1} & -
&  $335\pm52\pm47\pm12\pm25$
\\ \hline
\end{tabular}
\end{table}
 From the table, we see that our results of $f_{\eta_c}$
 from the LFQM I are smaller than those from the LFQM II due to the use of a smaller decay width of
 $\eta_c\to\gamma\gamma$. We note that 
 the center values of our  results on $f_{\eta_c}$ in Table~\ref{T1}
  correspond to  the use of $m_c=1.29$ GeV. 
 It is clear that in order to match the experimental result in the CLEO collaboration~\cite{etacdec1}
  and the Lattice QCD value~\cite{etacdec2}, a smaller mixing case with a larger $m_c$ is favored. 
  We emphasize that
 our predictions for $f_{\eta_c}$ are sensitive to
 the charm quark mass $m_c$.   
    In Fig.~\ref{F3}, we  show the charm quark mass dependence for the $\eta_c$ decay constant in the non-mixing case (c) 
    in Eq.~(\ref{Mixing}). 
 From the figure, we observe that to fit 
  the  CLEO data~\cite{etacdec1} or
the Lattice QCD value~\cite{etacdec2},
%
a large value of $m_c$ is needed.

\se{Conclusions}

We have studied the  transition  form factor of $\eta_{c}\to \gamma^*\gamma$ and the decay constant of 
 the $\eta_{c}$ meson in the LFQM. 
In particular, we have illustrated the transition form factor of  $\eta_c \to \gamma^* \gamma$ 
as a function of the momentum transfer $Q^2$.
We have shown that although our results are consistent  with the experimental data 
by the BaBar collaboration, they are about 10$\%$ smaller than the data points
for $Q^2$ in the range of 7.5$\sim$20 GeV. 
We have also evaluated the decay constant of $\eta_c$. We have shown that it is sensitive to 
the mixing angles as well as 
the mass of the charm quark.
Explicitly, for $\eta_c\sim c\bar{c}$, i.e. a pure $c\bar{c}$ state ($\theta_c=0$),
we have found that
$f_{\eta_{c}}=230.5^{+52.2}_{-61.0}$
 and $303.6^{+115.2}_{-116.4}$   MeV 
in the LFQM I and II based on the two sets of input parameters, $\Gamma_{\eta_c \to \gamma \gamma}=5.3\pm 0.5$
and $7.2\pm2.1$ keV, given by  Particle Data Group and Lattice QCD calculation, respectively.
Both results are within the error of $335\pm52\pm47\pm12\pm25$ MeV measured by
the CLEO collaboration, but they are somewhat  smaller than $394.7\pm2.4$ MeV predicted by the Lattice QCD,
in which $\Gamma_{\eta_c\to \gamma\gamma}=7.2\pm2.1$ keV is used like the LFQM II.
However, the Lattice QCD result can easily be accounted when a larger value of the charm quark mass is used.
Future precision measurements on the decay width of $\eta_c\to \gamma\gamma$ are clearly needed in order 
to determine the $\eta_c$ decay constant in the LFQM.

\begin{acknowledgments}
 This work was partially supported by National Center for Theoretical
Sciences, SZL-10004008, National Science Council  
 (NSC-97-2112-M-471-002-MY3, NSC-98-2112-M-007-008-MY3, and
 NSC-101-2112-M-007-006-MY3) and National Tsing-Hua
University (102N1087E1 and 102N2725E1)
\end{acknowledgments}

\end{document}